\documentclass[12pt]{article}
\usepackage{amsmath}
\usepackage{xcolor}
\usepackage{graphicx,psfrag,epsf}
\usepackage{comment}
\usepackage{enumerate}
\usepackage{booktabs}
\usepackage{threeparttable}
\usepackage{adjustbox}
\usepackage{tabu}
\usepackage{natbib}
\setcitestyle{aysep={}}

\usepackage{url} % not crucial - just used below for the URL 
\usepackage{array}
\newcolumntype{P}[1]{>{\raggedright\arraybackslash}p{#1}}
\usepackage[font=small]{caption}

\usepackage{setspace} % for appendix spacing
\usepackage{geometry} % for appendix spacing
\usepackage{url} % Required for url
\usepackage{hyperref} % Required for linking; apparently this is typically the last package to be used
\usepackage{placeins} % possibly needed for floatbarrier

\newcolumntype{d}{S[table-format=3.2]}

\newcommand{\blind}{0}

\begin{document}

\def\spacingset#1{\renewcommand{\baselinestretch}%
{#1}\small\normalsize} \spacingset{1}

%%%%%%%%%%%%%%%%%%%%%%%%%%%%%%%%%%%%%%%%%%%%%%%%%%%%%%%%%%%%%%%%%%%%%%%%%%%%%%

\if0\blind
{
  \title{\bf The Software Behind the Stats: A Student Exploration of Software Trends in Economics, Political Science, and Statistics}
  \author{Elizabeth Upton (contact author)\\
  emu1@williams.edu\\
    Department of Statistics, Williams College\\
    18 Hoxsey St.\\
    Williamstown MA 01267 \\
    $~$ \\
    Xizhen Cai \\
    xc2@williams.edu\\
    Department of Statistics, Williams College\\
    $~$ \\
    Pamela Jakiela \\
    pj5@williams.edu\\
    Department of Economics, Williams College\\
    $~$ \\
    Owen Ozier \\
    oo3@williams.edu\\
    Department of Economics, Williams College\\
    $~$ \\
    Shyam Raman \\
    sr21@williams.edu\\
    Department of Economics, Williams College}
  \maketitle
} \fi

\if1\blind
{
  \bigskip
  \bigskip
  \bigskip
  \begin{center}
    {\bf The Software Behind the Stats: A Student Exploration of Software Trends in Economics, Political Science, and Statistics}
\end{center}
  \medskip
} \fi

\newpage
\begin{abstract}
This paper presents a student activity designed to explore the use of software and reproducibility practices in academic research across selected journals in economics, political science, and statistics. The activity was designed to deepen students’ understanding of reproducible research workflows and increase their awareness of disciplinary norms in quantitative scholarship. Students reviewed reproduction files associated with published articles from major journals and repositories, using a survey to document software use. Combined with web-scraped metadata, the resulting dataset spans more than 10,000 papers and reveals clear patterns within the journals examined: Stata remains dominant in economics, R is increasingly common in political science, and R is the standard in statistics, with a growing number of articles using multiple software platforms within a single manuscript. These results provide educators with a data-driven view of software practices in contemporary quantitative research, helping to inform decisions about which tools to emphasize in the classroom. To assess whether the student activity met its pedagogical goals, we summarize student feedback, which indicates increased understanding of academic workflows and greater awareness of software diversity. We provide all instructional materials, data, and code to support adoption across course levels and disciplines, and conclude with suggestions for follow-up assignments. 
\end{abstract}

\noindent%
\vfill

\newpage
\spacingset{1.45} % DON'T change the spacing!
\section{Introduction}
\label{sec:intro}
A range of academic departments, including statistics, economics, and political science, offer courses in statistical and econometric methods.  The teaching is not entirely theoretical; recent work has documented and encouraged a pedagogical shift towards empirical examples and data-centric projects  \citep{AngristPischke2017undergraduate, gaise, ws}. Classes on these topics now regularly require students to work with widely-used statistical software packages.  Which programming language or statistical package, however, seems to vary across disciplines and is constantly changing over time \citep{Bednarowska2025DataScienceClassroomChallenges}.

% OO: OLD SENTENCE: Classes on these topics now regularly include exercises that require the use of widely-used statistical software packages.
% OO: NEW SENTENCE: Classes on these topics now regularly require  students to work with widely-used statistical software packages.

As educators, we make a conscious decision about what software tools we will use in our courses \citep{Schwab2021datascience,Siegfried2021trends}.  This choice is often based on our own experiences, backgrounds, and the resources available to us.  A pragmatic consideration should also be the current state of the software landscape---that is, an understanding of what software is being used in different disciplines.  For example, economics departments teach in Stata at least in part because published economic research uses Stata---or at least, conventional wisdom suggests that this is the case, though data on this front is limited.

From the student perspective, understanding the software landscape is also valuable. Students make decisions about which courses to take and how much effort to invest in learning a particular software tool or programming language. This decision may be particularly salient with the rapid adoption of generative AI tools, which can produce code and improve immediate task performance without necessarily producing corresponding gains in learning or longer-term knowledge transfer \citep{dan}. Developing fluency in a tool requires time and motivation. The classic question—``why do I need to know this?"—can partially be answered with a clearer picture of disciplinary software norms, which conveys why particular tools are worth learning and how they connect to authentic research practices~\citep{deveaux}.

While software preferences shift, one fundamental principle remains constant: reproducibility (i.e. obtaining consistent results using the same input data, computational steps, methods, code, and conditions of analysis \citep{NAP}) is essential for rigorous research 
\citep{Munafo2017manifesto,GertlerGalianiRomero2018Nature,Vilhuber2022teaching,WrobelMcGowan2024JASA, reproduce}.  Yet, while we tell our students that their code should be shareable and readable, they may not understand how reproducibility pipelines are operationalized in practice.  Students may initially be unaware of the idea of reproducibility in data analyses \citep{Ostblom2022JSDSEopinionated}; even in classes that use data, a recent survey showed that relatively few classes introduce students to common ways of transparently sharing data and code \citep{Underwood2024JSDSEteaching}. 

Additionally, we often do not emphasize how reproducible workflows can be leveraged by students in their own research.
That is, reproducibility is not just about making their own work transparent for others—its existence enables students to reproduce findings from a paper, learn how to practically apply methods, and build upon existing research with confidence~\citep{ball, xiong2023state}.  Our courses, then, should not only teach students how to use the statistical tools prevalent in their field but also foster an understanding of reproducibility as a core research practice~\citep{horton}.  At a low cost, we could also expose students to the tools used in other disciplines, emphasizing how the skills developed in one software package can transfer to another—and how these tools, in combination, support reproducible research.

To this end, we designed a short activity for students that provides hands-on practice in accessing and interacting with existing reproduction packages, an essential first step in fostering reproducible research skills. The results of the activity, combined with web-scraped metadata, form a dataset describing software use in over 10,000 academic publications across statistics, economics, and political science journals. This dataset offers insight into the current landscape of statistical software usage, information that is valuable for students interested in quantitative fields of study and for instructors making decisions about which software tools to use in their classes. Thus, the contributions of this paper include (1) the introduction of a student activity, easily adaptable to a variety of classroom settings, that exposes students to reproducibility in academia; (2) the use of the resulting dataset to identify software trends that can inform learning and teaching practices; and (3) an outline of several potential extensions that build on or make use of the dataset, including student activities focused on reproducibility verification, documentation quality, and software comparison.

The remainder of the manuscript is organized as follows. We first provide historical context and review prior work on trends in software use in research and teaching. We then describe the data sources and the student activity used to construct the dataset, present findings derived from the dataset, and conclude with a discussion of student outcomes and proposed extensions.  Throughout the manuscript, we distinguish between reproducibility and replicability (obtaining consistent findings when a study is repeated with new data addressing the same scientific question) following~\cite{NAP}. We note, however, that these terms have often been used interchangeably in the literature, including in some of the articles reviewed and in the original design of the student activity.  

This study was reviewed by the Williams College Institutional Review Board and determined to be exempt from IRB review, as it presents no risk to participants and relies exclusively on publicly available data.  To support transparency and reproducibility, all data and analysis code used in this study are available in an Open Science Framework (OSF) repository: \url{https://osf.io/72ajq/}. A README file provides an overview of the repository and identifies the data and R scripts used to generate each figure in the manuscript. R scripts in the Code folder are named according to the figures they produce (e.g., fig-1-...).

\section{Background}

To contextualize the changing software landscape, approximate release dates for selected software packages and programming languages are provided in Table \ref{tab:timeline}. Although many additional entries could be included, the table is not intended to provide a comprehensive history; rather, it places today's most widely used tools in historical perspective. Statistical software options have changed dramatically over recent decades, as have computing technologies more broadly. Despite these changes, there has been relatively little comprehensive documentation of trends in statistical software use in academic publications. We discuss some recent reviews below, although each has its own limitations.
% \cite{Renfro2004compendium} describes the history of related software in much more detail, including the development of TSP on Univac machines, the transformation of MicroTSP into EViews, and so on.}
%In addition to statistical packages, the table includes some programming languages (such as Python and Julia),
%In addition to statistical packages and interactive environments, the table includes two programming languages---Python and Julia---since some research and teaching may use them.
%\footnote{Older languages such as FORTRAN and C are still used directly in some research, but we omit them from the table as we are not trying to establish a complete history of mathematical computing.}  

% note appendix can be found at SSRN version
% https://papers.ssrn.com/sol3/papers.cfm?abstract_id=4620006
%In terms of how statistical software is used in practice, there are some recent reviews,
\cite{FivsarGreinerEtAl2024ManagementScience}, for example, report on relative software popularity for articles published in \emph{Management Science} between 2019 and 2023.  They find that Stata is the most commonly used (60\% of articles), followed by R (19\%), MATLAB (18\%), SAS (13\%), and Python (11\%), with other analytical software being used much less frequently.  This provides a recent picture, but at one journal and for only a few years.

\cite{Vilhuber2020reproducibility} documents trends from 2010 to 2019, restricting attention to journals published by the American Economic Association.  He shows that Stata is present in the majority of reproduction packages for those journals in every year, with MATLAB in second place, and a changing cast of others in a relatively distant third (SAS in some years, R or other software in others).

\cite{christodoulou2015stata}, in an analysis published by Stata Press, provides some earlier patterns of software use. He describes a literature search procedure for ``citations of notable statistical software: GenStat, GLIM, MATLAB, Minitab, NLOGIT, Octave, RATS, SAS, SHAZAM, SPSS, Stata, Systat, and TSP.''  The analysis excludes the ubiquitous  Microsoft Excel because ``it is not marketed as a statistical software,'' but makes a more particular note regarding the approach's limitation in relation to R:  ``The most notable omission is the R-project due to the commonality in its keyword (that is, the single letter `R').'' (p.101)
%Our approach, enabled by the practice of making reproduction files accessible, circumvents this problem.  
That analysis shows that as far back as 2000, in each of the research areas  described as ``Business,'' ``Accounting,'' and ``Finance,'' the leading two statistical software packages were consistently SAS and SPSS.  By 2013, in all three areas, SAS had dropped out of the top two spots, and Stata was one of the top two, along with MATLAB or SPSS.  Relative citation counts from 1999 to 2013 are shown in that study's Figures 13.1, 13.2, and 13.3.  The graphs include SPSS, Stata, SAS, MATLAB, RATS, EViews, S-Plus, and Limdep.

% The top positions at three points in these graphs are described in the table below (with a more-than-two-way tie simply listed as ``tie''):

% \begin{tabular}{c|ccc|ccc|ccc|}
%      Field:
%      & \multicolumn{3}{c|}{Accounting} 
%      & \multicolumn{3}{c|}{Business}
%      & \multicolumn{3}{c|}{Finance} \\
%      Year:
%      & 1999 & 2006 & 2013
%      & 1999 & 2006 & 2013
%      & 1999 & 2006 & 2013 \\
%      \hline
%      Rank 1 
%      & SAS & SAS,SPSS & SPSS
%      & SPSS & SPSS & SPSS
%      & RATS & Stata & Stata
%      \\
%      Rank 2 
%      & SPSS &  & Stata
%      & SAS & SAS & Stata
%      & SAS & SAS & Matlab
%      \\
%      Rank 3
%      & (tie) & Stata & SAS
%      & Stata & Stata & SAS
%      & SPSS & Matlab & SAS
%      \\
% \end{tabular}

\begin{table}
%\caption{Selected statistical and econometric software release timeline.  \label{tab:timeline}}
\caption{Selected statistical computing release timeline.  \label{tab:timeline}}
\begin{center}    
\begin{tabular}{c|ll}
     \textbf{Year} & \textbf{Software} & \textbf{Citation} \\
     \addlinespace[1em]
     \hline
     \addlinespace[1em]

     1968     &  SPSS  & \citep{WilsonLorenz2015} \\
     1972     &  Minitab & \citep{Schissler2019statistical} \\
     1976     &  SAS  & \citep{Rodriguez2011sas} \\
     1976     &  S  & \citep{Chambers2020} \\
     1980     &  RATS  & \citep{Renfro2004compendium} \\
     1981     &  LimDep  & \citep{Renfro2004compendium} \\     
     1984     &  MATLAB  & \citep{ChonackyWinch2005Matlab} \\
     1984     &  GAUSS  & \citep{Anderson1992GAUSS} \\
     1985     &  Stata  & \citep{Renfro2004compendium} \\
     1991     &  S-Plus  & \citep{Hallman1993SPlus} \\
     1991     &  Python  & \citep{VanRossumDrake2003Intro} \\
     2000     &  R  & \citep{Chambers2020} \\
%     1995/2000     &  R  & \citep{Schissler2014statistical,Chambers2020} \\
     2012     &  Julia  & \citep{Perkel2019julia}
\end{tabular}
\end{center}
\end{table}

%\citep{Pinzon2015thirty} had been cited but is replaced by \citep{Renfro2004compendium}  in the table and by \citep{christodoulou2015stata} in the trends discussion.

We found comparatively little research examining trends in software use over time in undergraduate teaching. Instead, much of the existing literature focuses on qualitative considerations such as the relative cost or ease of use of different software tools. For example, \cite{ConawayClarkAriasFolk2018integrating} note that “students do not seem to find R as easy to learn as Stata,” while also emphasizing that R may be the most practical option for students without access to proprietary software.

In the absence of systematic studies documenting trends, textbooks offer an alternative lens for understanding the instructional software landscape. Some econometrics texts emphasize a single language—for instance, \cite{AngristPischke2009MHE} includes Stata code throughout—while others reflect shifts over time. The econometrics text by \cite{wooldridge2016introductory}, for example, long provided datasets in formats such as Stata, Eviews, Minitab, and Excel, but began including R datasets in its sixth edition (2016). More recent texts often adopt a multi-language approach: \cite{Bailey2019real} includes both R and Stata examples, explicitly discussing trade-offs between usability and cost.  In the field of statistics education, one can find teaching examples in textbooks that use R, Python, SPSS, Stata and more \citep[e.g.,][]{Ferrall1995JSE,Tucker2023JSDSE, de2005stats}. The data science text by \cite{baumer2017modern}, utilized across disciplines, uses R throughout, while acknowledging Python as a widely used alternative and noting the lack of consensus around a single “best” language.

\begin{figure}
\begin{center}
\vspace{-10pt}
\includegraphics[width=0.9\textwidth]{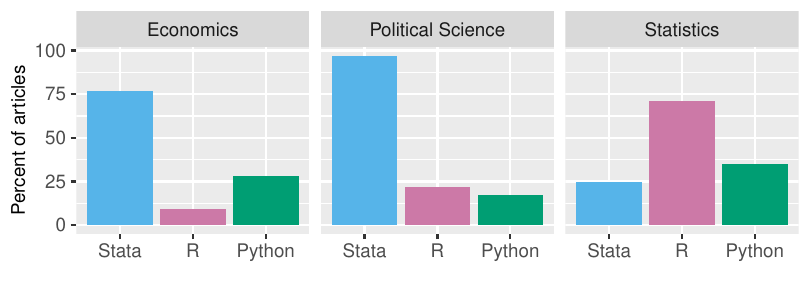}
\vspace{-5pt}
{\small
\caption{Mentions of software in three journals.  Percents are reported among articles mentioning any of ``Stata'' or ``Python'' or ``Rstudio'' or ``R programming'' or ``R code'' in the \emph{Journal of Economic Education}, the \emph{Journal of Political Science Education}, and the \emph{Journal of Statistics and Data Science Education}, from 1995 through 2024.  Sample sizes: 43 in economics; 23 in political science; 154 in statistics and data science.} \label{fig:threeeducationjournals}}
\end{center}
\end{figure}

There do appear to be disciplinary differences in software choice. To illustrate this, we conducted keyword searches for articles published over the past 30 years (1995–2024) in journals focused on teaching within each of three disciplines (Figure~\ref{fig:threeeducationjournals}). For Stata and Python, we search for a single keyword.  For R, we search for  ``Rstudio'' or ``R programming'' or ``R code'' rather than R, aware this may under count examples of R. The comparison of relative rates across journals is still informative even if the levels suffer from differential biases.  

In both the \emph{Journal of Economic Education} and the \emph{Journal of Political Science Education}, more articles explicitly mention Stata than mention either Python or R. However, in the \emph{Journal of Statistics and Data Science Education}, the pattern is very different: there are more articles mentioning R or Python than there are mentioning Stata. A recent article by \cite{BestMallinson2024quantitative} describes a shift over the last two decades in quantitative political science away from ``proprietary software like SPSS, Minitab, SAS, or STATA'' towards a ``more diverse'' set of tools including ``R and Python.''  This summary of pedagogy journals is limited, however: we examine only a single journal per discipline, which may or may not be representative of the entire landscape.  Furthermore, with only 43 total articles mentioning any of these types of software in the \emph{Journal of Economic Education} and only 23 such articles in the \emph{Journal of Political Science Education}, we cannot say much about trends from this dataset.

%\footnote{Historically, this journal primarily focused on statistics.  In 2021, it incorporated ``Data Science" into its name to reflect the growing importance of the field.} 

\section{Data Collection and Student Activity}
%\section{The Activity}
To characterize software trends over time, and to get a more detailed look at software use than the simple illustration in the previous section provides, we built a large dataset of published papers.
% OO: new transition sentence above seemed like it could help.
Our dataset documenting software usage across statistics, economics, and political science was constructed using a two-pronged approach. First, we systematically scraped large online repositories to collect repository metadata and file extensions. The code used for this step is available in the OSF repository under Code/webscraping. Second, via the student activity, students visited a sample of articles and their associated repositories to examine the available reproduction files directly. The two approaches complement one another. The automated approach collects substantially more data in far less time, while the manual review is particularly valuable for journals that do not maintain a centralized repository for reproduction files. Even when automated collection is possible, manual review validates the scraped data, while the scraped data provide an independent check on the completeness and consistency of the manual review. We elaborate on both approaches below.  

\subsection{Data Sources}
\label{sec:data}

The five major sources of data include three specific journals or publishers: the \emph{Journal of the American Statistical Association} (JASA), the \emph{American Economic Review}, and the \emph{American Political Science Review}.  These are each flagship journals of the scholarly association associated with each field in the United States.
Two other data sources are the 
repository for the American Economic Association hosted by the Inter-university Consortium for Political and Social Research (ICPSR) and the Harvard Dataverse---which includes analysis repositories from multiple journals in multiple disciplines, and for which we focus on economics and political science journals.    Across the Harvard Dataverse and the American Economic Association repositories at the ICPSR, our automated data collection includes all the so-called ``Big 3'' journals in political science, as well as three of the so-called ``Top 5'' journals in economics. Another reason to use the American Economic Association repository is that that organization's Data Editor has supported the curation of organized and quality-checked programs and data files supporting research articles for several years \citep{hoynes2023reproducibility}.  
Details on the sample sizes, journals, and years covered by these sources are provided in Online Appendix Table A1.  Note that the data were scraped in the fall of 2024, affecting 2024 sample sizes for the automated data collections.

%\footnote{The American Economic Association collection at ICPSR has been studied in other ways as well; for example, by \cite{Li2024crowdsourcing}. \cite{Underwood2024JSDSEteaching} have also highlighted the Dataverse as a leading example of  research accessibility resources from which students might benefit.}

Sources were chosen based on our familiarity with them as scholars, their topical relevance and comprehensiveness, and the ease of accessing information.  For example, we explored including articles from the Institute of Mathematical Statistics (i.e.~\emph{Annals of Statistics} and \emph{Annals of Applied Statistics}) but found it difficult to get a succinct list of DOI stubs and for students to be able to reliably navigate to article content.  These choices necessarily limit the scope of our conclusions:  our findings should not be generalized to all of statistics, economics, or political science. We return to this limitation in the discussion, where it becomes a productive point of reflection for students.

\subsection{Student Assignment Overview}
\label{sec:assignment}
%We adopted a two-pronged approach to collecting data on software popularity within the fields of economics, political science, and statistics. A detailed description of our data sources can be found in Section~\ref{sec:data}. Briefly, some sources provided article-level metadata that could be automatically scraped, while others required manual review.

%For the latter,
For the manual approach to data collection, we developed a student assignment in which participants visited the webpages of journal articles. This gave students an opportunity to engage directly with reproduction materials as they appear in practice. Each student reviewed a small number of articles—across economics, political science, and statistics—identifying the software used. This familiarized students with reproduction practices in three fields, while enabling us to build new datasets and to validate our automated analysis.
%For statistics, we then aggregated their contributions to build a new dataset.  For economics and political science, we used the activity mainly to familiarize students with reproduction practices in those fields and to validate the student work and automated approaches against one another.  

Students participating in the activity received an assignment (see supplementary materials) with the prompt: 

\emph{``The goal is to see how researchers working in different disciplines share materials for reproduction of empirical statistical analyses (and what software they use). We will visualize the results once everyone has done their small piece of data collection. For your part, this means: (a) visiting a website relating to a published paper, (b) exploring it in a structured way, (c) putting what you find into an online survey, and (d) recording in our course website that you have done so.''}  

The students were instructed on how to access their assigned websites, either via a provided URL or a shortened digital object identifier (DOI). Their instructions included guidance on how to identify relevant files within reproduction packages and repositories: we gave them a list of file types to look out for (e.g. \textbf{.R, .Rda, .do, .py}) and instructions for how to open compressed folders. If no reproduction files were available, students were asked to skim the article for terms indicating the software used or alternative locations where the files might be found.

Once students completed their search, they recorded their findings in a structured Qualtrics survey. The survey began with general information collection:
\begin{enumerate}
    \item Provide the URL or DOI stub associated with this survey entry
    \item Before you answer detailed questions about this entry, was there anything unusual that you would like to comment on?
    \item Were you able to find any reproduction files associated with this DOI? (choice: Yes/No)
\end{enumerate}
If the student selected yes on question 3, they were presented with a series of follow up questions, such as:
\begin{itemize}
    \item Were you able to see any R code files (ending in .R, .Rda, .Rmd)
\end{itemize}
with response options Yes/No.  If the student selected no on question 3, they were asked if they found any description of the software used in the article.  An answer of no would bring them to the final survey question and an answer of yes would ask them to choose the software mentioned and provided a textbox for them to enter the exact wording used by the authors in the article.  The final survey question asked students whether they considered categorizing the files as a clear or ambiguous task.  To read the complete Qualtrics survey blueprint, see the supplementary materials.  

To ensure familiarity with the process before collecting data, we prepared two warm-up exercises for the students to complete.  We requested that they complete the survey based on fictional entries, which we describe for them in detail.  We also provided the students with a list of example DOIs whose content was known.  In preparation for their own data collection, they could visit these links, explore the files, and verify the types of files they identified against the known content.

\subsection{Student Assignment Protocol and Sample Sizes}

Five professors at the authors' college incorporated the assignment into their courses, three from economics and two from statistics. Before introducing the activity in classes, we conducted a test run with a small group of research and teaching assistants. These students also provided feedback on the average time required to complete the task.  All agreed that to check ten articles or reproduction packages via DOI took less than two hours; for all but one student it took less than one hour.  We confirmed the duration by asking student participants how long the larger-scale activity took them.  A typical duration reported for ten such checks was one hour: 24.5\% said it took between 20 and 40 minutes; the plurality (43.9\%) said it took between 40 minutes and 1 hour; 28.6\% said it took between 1 and 2 hours, and 3.1\% said it took more than 2 hours. 

In total, 158 students across seven separate classes and sections were assigned to review DOIs.  Each student was assigned four DOIs from JASA and six DOIs from economics and political science.
%(four per student, with each DOI assigned to two students); for pedagogical and data quality purposes, an additional 474 DOIs (six per student) were from sources in economics and political science, for a total of 790.
To assign DOIs, we created a master file of student IDs and randomly allocated 10 DOI stubs to each student. Each article was initially assigned to two students for review, although never to two students in the same class. This resulted in 790 unique DOIs across all journals.

As an example, 316 JASA DOIs were initially assigned. Of these, 214 were visited by at least one student and yielded a usable classification outcome. Among the 214 classified DOIs, 131 were independently reviewed by multiple students who agreed on every survey question, 9 were reviewed by at least three students with a majority agreement on classification, and 74 were reviewed by a single student. Fourteen assigned DOIs were never visited because some students did not complete the assignment. An additional 88 DOIs were reviewed by an even number of students (almost always two) who disagreed on at least one survey question; these cases were excluded from the analysis.  Of the 214 classified DOIs, 69 did not have any files to report or describe because they corresponded to editorials, theoretical or review articles, or articles for which no reproduction files were available. The remaining 145 JASA articles form the basis of the analysis presented in Figure \ref{fig:2024jasa}.

Interested students who had asked for more work of this kind were hired as research assistants.  They checked some of the DOIs that previous students had missed or that had been inconsistently classified previously, and they carried out additional comprehensive data collection for the most recent years of three journals.

\subsection{Benefits of the Two-Pronged Approach}

To cross-check automated collection and to verify student data quality, we assigned a small number of automatically classified Dataverse entries to students.  Of 68 repositories of this kind, we checked MATLAB, Stata, and R status across the automatic and student classification, for 204 comparisons.  There were 201 agreements and 3 disagreements.  In one case, the students correctly noticed a filetype that our initial automated procedure had incorrectly omitted.  We revised the automatic procedure to capture this.  In the other two cases, the students reported seeing both Stata and R when in fact there was only R.  The student error rate is thus roughly 1 percent.

As mentioned, there are limitations to the scrapable metadata that can be overcome by our student activity.  For JASA, the manual student approach produces more current data than is otherwise available.  We also note that the website of the American Economic Association, in its Terms of Use, forbids the use of ``scraping programs'' to harvest content directly from its site.  For the Dataverse, the manual student-driven approach allows us to cross-check our use of metadata, and allows us to know what is inside zipped archives that are otherwise not transparent via metadata.  Furthermore, conditioning on the existence of an online repository does not allow us to answer the question, “What fraction of papers are accompanied by reproduction files?” Repository scraping necessarily restricts attention to manuscripts that already have repositories; papers without reproduction materials are never observed and therefore cannot be counted. In contrast, student inspection at the journal level allows us to estimate the proportion of all manuscripts within a journal that include reproduction files, and also captures papers that link to external data or code sources outside standard repositories. 

%\footnote{There is a set of 109 GitHub repositories associated with JASA from 2018 to 2021, but we do not graph the trends there in detail both because it is a smaller dataset than any of the others reported in this paper, both because it is not clear whether it is a representative sample from that time period
%and because that series does not extend to the present.  We do note that 83\% of those repositories use R; 16\% use MATLAB; 17\% use C or C++; 3\% use Julia, 2\% use Python, and 1\% use each of Stata, Stan, and FORTRAN.}  

\begin{comment}
which students did for all articles in the most recent year of \emph{Journal of the American Statistical Association}, the \emph{American Political Science Review}, and the American Economic Review (one of the AEA journals), along with a sampling of articles from earlier years.
\end{comment}

Both the automated analysis of file types and our student-driven activity examining files themselves overcome the limitations of purely keyword-based analysis mentioned in previous work \citep{christodoulou2015stata}.

%We then summarized the findings from student work on \emph{Journal of the American Statistical Association}, together with automated approaches for other data sources, and  shared the results with our students.    
%
%Interested students who had asked for more work of this kind checked on some of the DOIs that previous students had missed or that had been inconsistently classified previously, and DID ADDITIONAL THINGS HERE WE SAY SOMETHING IF WE LIKE ABOUT THE THREE UPSETS.  

\section{Findings}

\begin{figure}
    \begin{center}
    \caption{Software in economics journal publications posted to Harvard Dataverse}
    \label{fig:econdataverse}
    \includegraphics[width=0.9\textwidth]{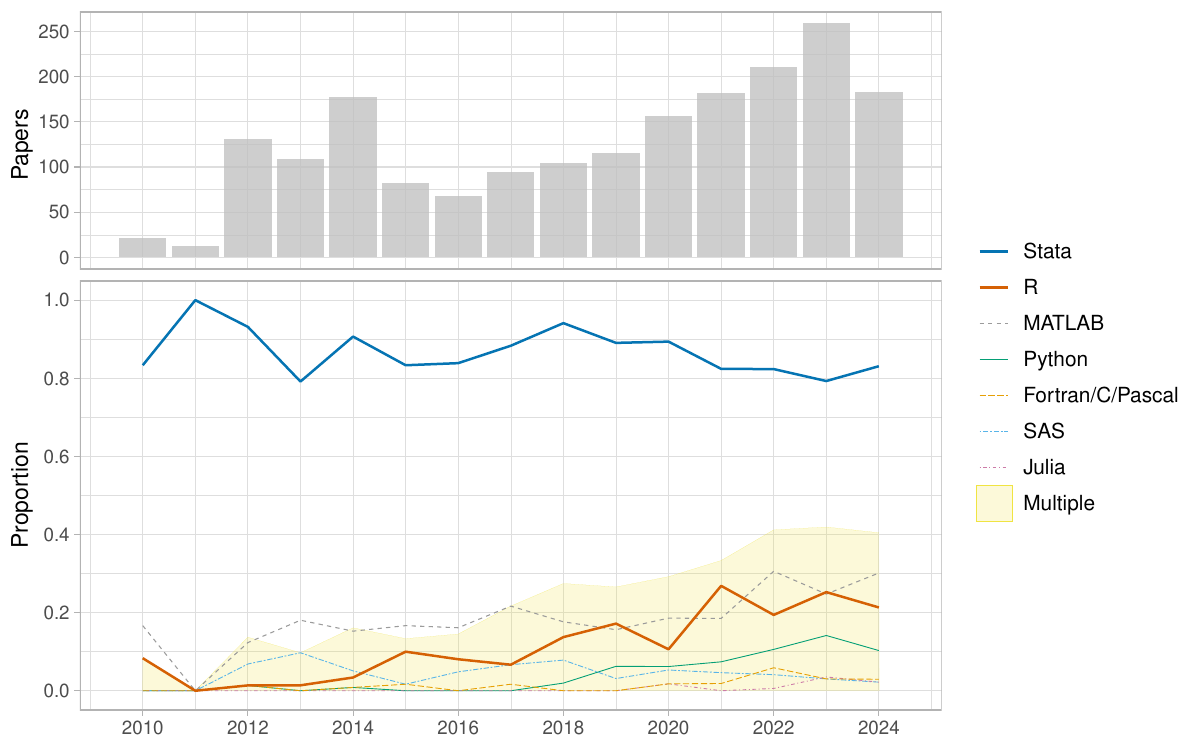}
    \begin{minipage}[c]{0.9\textwidth}\hangindent=1.75em
            {\footnotesize
            % Note goes here
                \textbf{Note:} Figure presents trends in software used by reproduction packages from $N=1{,}906$ papers published in economics journals. Data were collected via automated file categorization and validated via student exploration. Upper panel reports the number of articles examined in each year; lower panel reports proportions (scale from 0.0 to 1.0).

            % leave paragraph gap to reset hanging indent
            
            % Data source goes here
                \textbf{Source:} Data sourced from Harvard Dataverse for the following journals: \emph{Journal of Political Economy}, \emph{Journal of Political Economy: Macroeconomics}, \emph{Journal of Political Economy: Microeconomics}, \emph{Quarterly Journal of Economics}, and \emph{Review of Economics and Statistics}.
            } % close footnote size chunk
        \end{minipage}
    \end{center}
\end{figure}

\begin{figure} % fig env
    \begin{center} % center it
    \caption{Software in reproduction packages from the American Economic Association ICPSR repository.} % fig title
    \label{fig:icpsr} % fig label
    \includegraphics[width=0.9\textwidth]{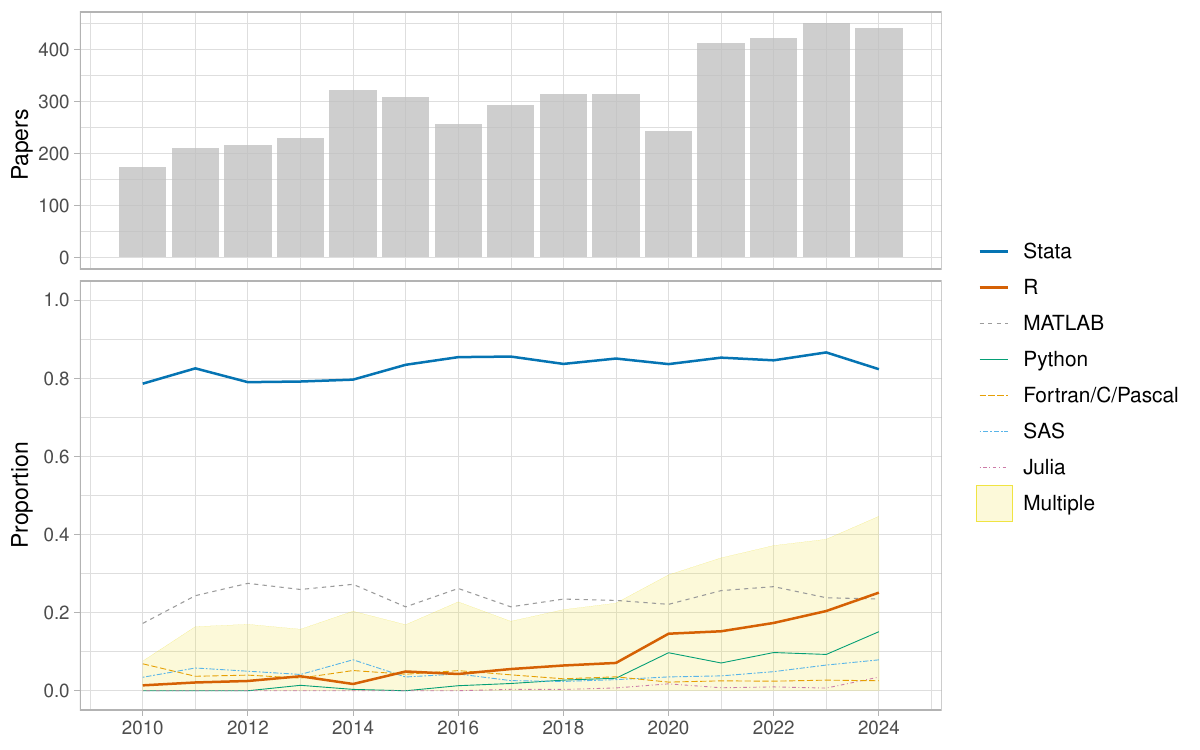} % figure
    \begin{minipage}[c]{0.9\textwidth}\hangindent=1.75em
        {\footnotesize 
        % Note goes here
            \textbf{Note:} Figure presents trends in software used by reproduction packages from $N=4{,}617$ papers published in economics journals. Data were collected via automated file categorization. Upper panel reports the number of articles examined in each year; lower panel reports proportions (scale from 0.0 to 1.0).
        
        % leave paragraph gap to reset hanging indent
            
        % Data source goes here
            \textbf{Source:} Data sourced from the American Economic Association's ICPSR repository for the following journals: \emph{American Economic Journal: Applied Economics}, \emph{American Economic Journal: Economic Policy}, \emph{American Economic Journal: Macroeconomics}, \emph{American Economic Journal: Microeconomics}, \emph{American Economic Review}, \emph{Journal of Economic Literature}, and \emph{Journal of Economic Perspectives}.
        } % close footnote size font chunk
    \end{minipage}
    \end{center}
\end{figure}

\begin{figure}
\begin{center}
\caption{Software in political science journal publications posted to Harvard Dataverse}
\label{fig:psdataverse}
\includegraphics[width=0.9\textwidth]{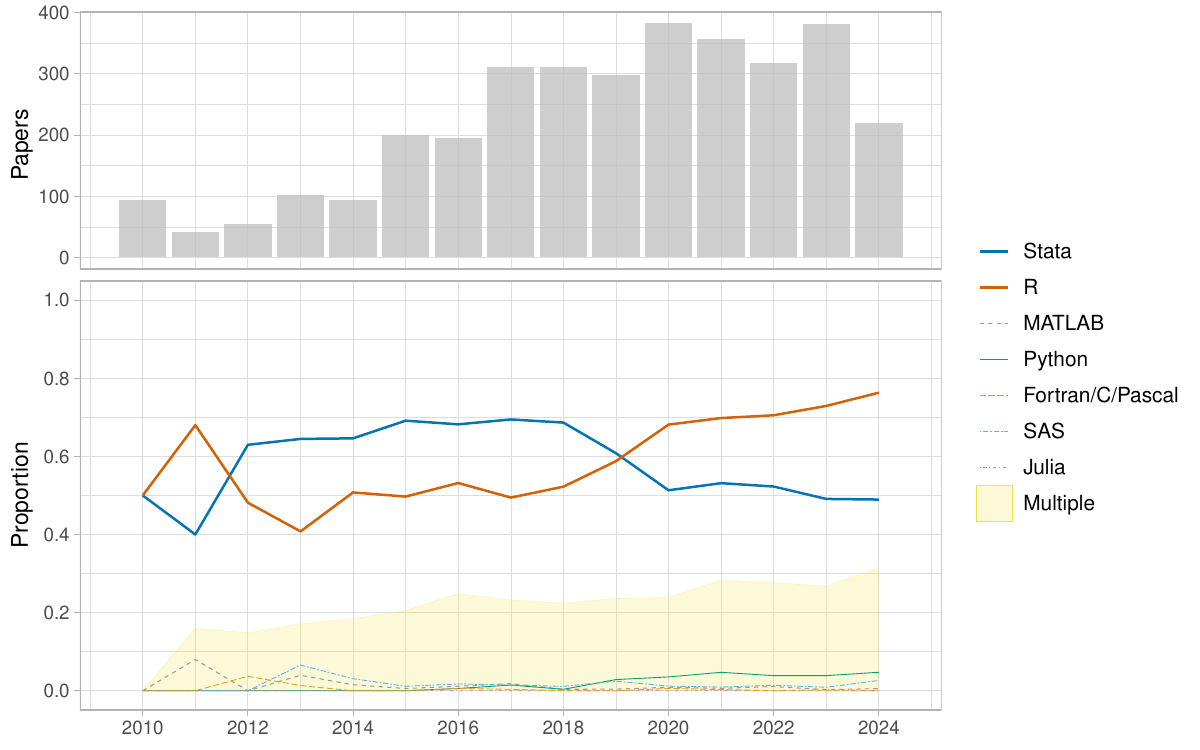}
    \begin{minipage}[c]{\textwidth}
        {\footnotesize
        % Note goes here
            \textbf{Note:} Figure presents trends in software used by reproduction packages from $N=3{,}363$ papers published in political science journals. Data were collected via automated file categorization and validated via student exploration. Upper panel reports the number of articles examined in each year; lower panel reports proportions (scale from 0.0 to 1.0).\\
        % Data source goes here
            \textbf{Source:} Data sourced from Harvard Dataverse for the following journals: \emph{American Journal of Political Science}, \emph{\emph{American Political Science Review}}, \emph{British Journal of Political Science}, \emph{Journal of Politics}, and \emph{Political Analysis}.
        } % close footnote size font chunk
    \end{minipage}
\end{center}
\end{figure}

\begin{figure}
\begin{center}
\caption{Trends in software for \emph{Journal of the American Statistical Association} articles}
\label{fig:jasatrend}
\includegraphics[width=0.9\textwidth]{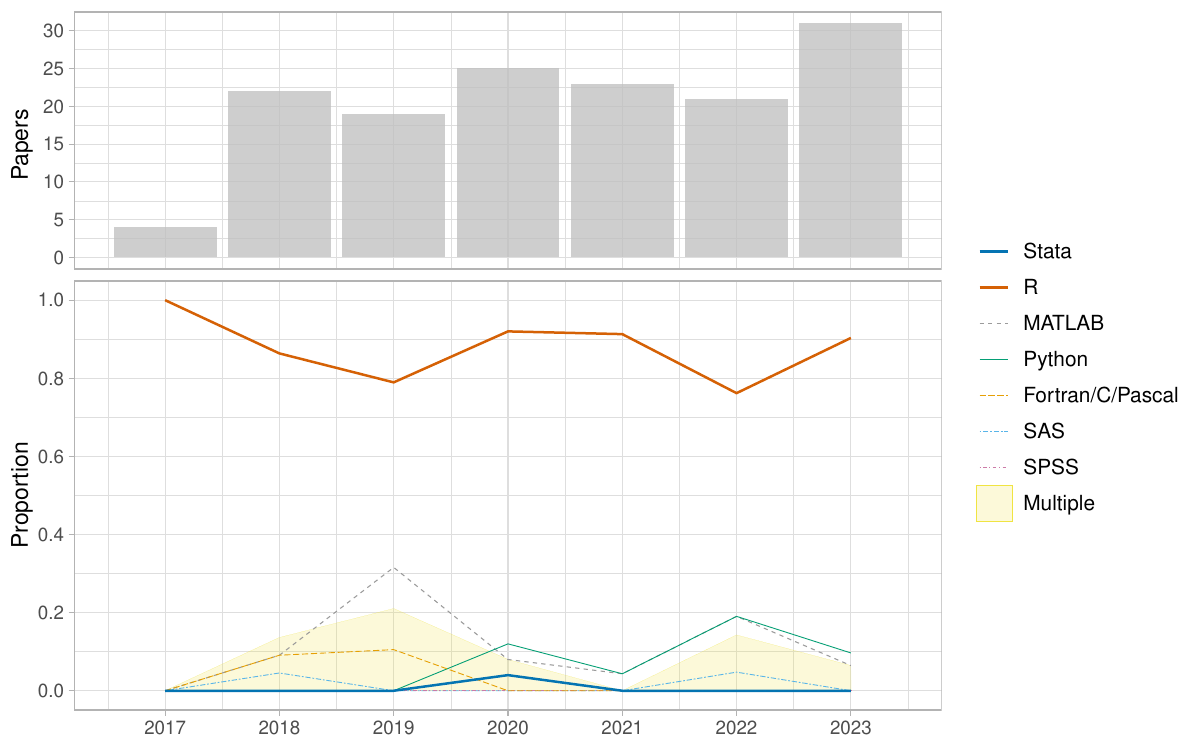}
    \begin{minipage}[c]{\textwidth}
        {\footnotesize
        % Note goes here
            \textbf{Note:} Figure presents trends in software used by reproduction packages from $N=145$ papers published in JASA. Data were collected via student exploration. Upper panel reports the number of articles examined in each year; lower panel reports proportions (scale from 0.0 to 1.0).\\
        % Data source goes here
            \textbf{Source:} Student data collection from the JASA website.
        } % close footnote size font chunk
    \end{minipage}
\end{center}
\end{figure}

Figures~\ref{fig:econdataverse} to \ref{fig:jasatrend} show statistical computing trends over time, as represented by our sample of published papers and reproduction packages from economics, political science, and statistics.

In the economics journals, Stata is dominant, appearing in 80\% (or more) of reproduction packages (Figures~\ref{fig:econdataverse} and \ref{fig:icpsr}). Moreover, there is no evidence that use of Stata is declining over time in economics. For much of the last fifteen years, MATLAB has been the second most widely used statistical software, but in recent years R has been catching up; in 2023 and 2024, both MATLAB and R were included in more than 20\% of reproduction packages in both the Harvard Dataverse and the American Economic Association/ICPSR data sets.  Python is also increasing in popularity: while almost no reproduction packages included Python files prior to 2016, Python is now present in more than 10\% of reproduction packages (and, again, this pattern is apparent in both the Dataverse and the American Economic Association/ICPSR data sets). 

As these numbers suggest, within economics there has been a general trend toward the use of multiple statistical software tools over the last fifteen years. In 2010, the proportion of reproduction packages that included two different types of statistical software was below 10\% in both the Dataverse and the American Economic Association/ICPSR samples. By 2024, more than 40\% of reproduction packages in both data sets included at least two different types of statistical software.  Thus, while Stata remains dominant in economics, the evidence suggests it is increasingly the case that economists use Stata in conjunction with other statistical computing tools such as MATLAB, R, and Python.

% Figure 2 -- ECON journal pubs available on Dataverse

% Figure 3 - ECON journal pubs available via ICPSR

In the political science journals, we also observe a broad trend toward the use of multiple statistical computing tools (Figure~\ref{fig:psdataverse}).  Both Stata and R are widely used within these journals, though R crossed an equal-popularity threshold in 2019 and appears to be trending in the positive direction, while Stata appears to be trending down slightly. However, Stata files are still present in approximately half of all reproduction packages from 2024, while R is now present in approximately 75\%. While the frequency of reproduction packages containing multiple types of statistical software has increased over time, the trend is not quite as marked as in economics: by 2012, more than 15\% of reproduction packages in our political science sample included at least two types of statistical software, and this increased to about 30\% by 2024. In contrast to economics, there is little evidence of MATLAB or Python (or any other type of statistical software) gaining ground in political science. 

While the broad trend in the economics and political science journals we reviewed appears to be toward greater use of multiple flavors of statistical software, there is little evidence of such a trend in the flagship statistics journal. Very few statistics articles (less than 1\%) in JASA (out of hundreds we checked) report using Stata or include Stata in their reproduction package (Figure~\ref{fig:jasatrend}).  Around 87\% of JASA articles use R, with MATLAB coming in a distant second (12\% of articles sampled) and Python third (8\% of articles).  There is some evidence that Python use is increasing over time, though overall adoption remains low. However, the evidence from statistics stands in stark contrast with what we observed in quantitative social science: Fewer than 10\% of JASA articles use more than one type of statistical software, and there is no evidence that this practice is increasing over time.

% Figure 4 - political scientists shifting from Stata to R, or 

% JASA random sample

\begin{figure}
\begin{center}
\includegraphics[width=0.8\textwidth]{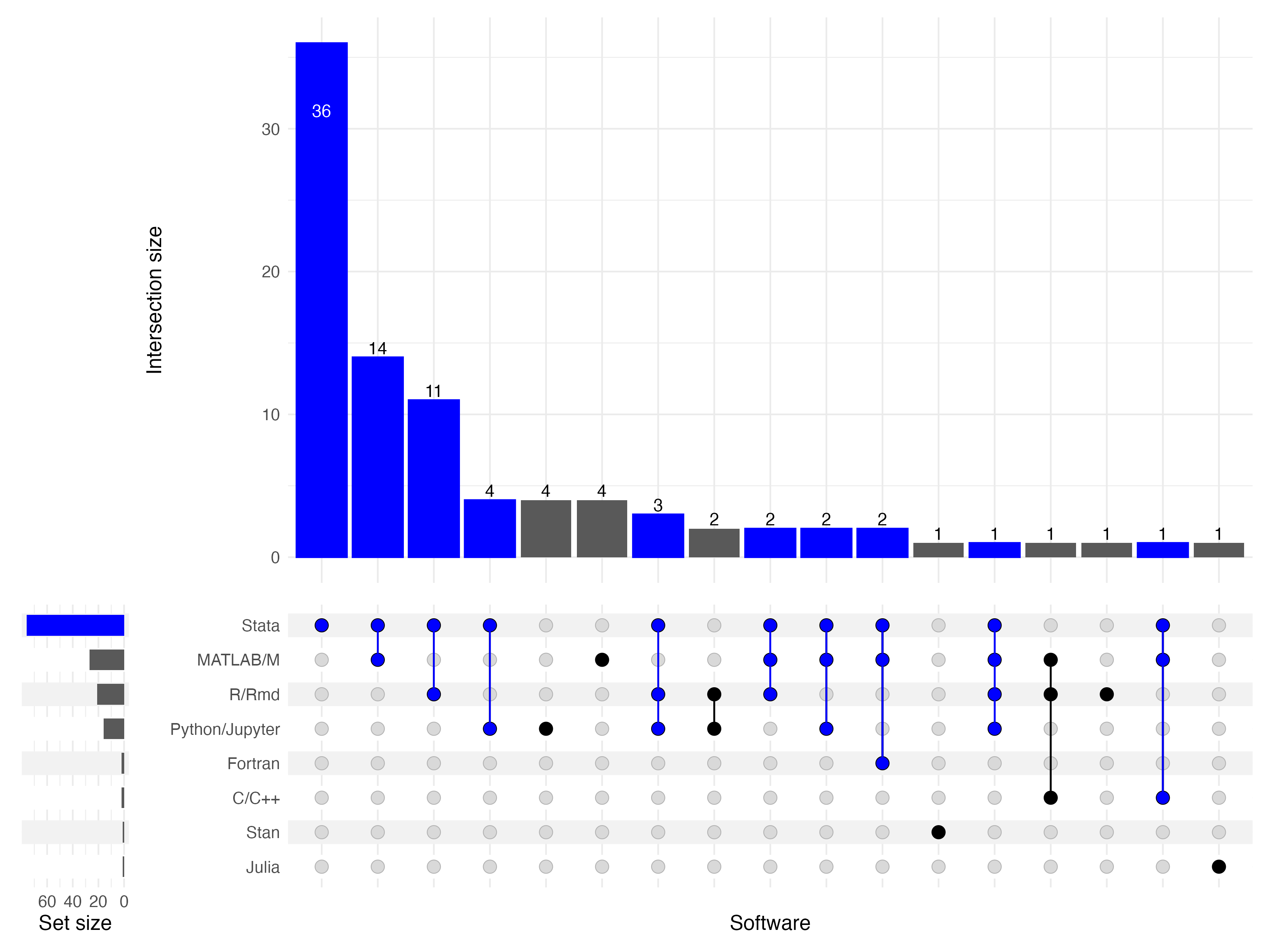}
\end{center}
\caption{Software in 2024 issues of the American Economic Review. Restricted to 90 articles (82 percent) in which software could be categorized out of 110 articles in total. \label{fig:2024aer}}
\end{figure}

To explore the increasing use of multiple software platforms and languages, we visualized the intersections of the tools in a single journal from each discipline in Figures~\ref{fig:2024aer}:\ref{fig:2024jasa}.  The data shown comes from 2024 and was collected via student exploration.  We see, for example, that 35 articles published by the \emph{American Economic Review} use Stata, 14 use Stata and MATLAB, and 11 use Stata and R.  In the \emph{\emph{American Political Science Review}}, R is used on its own in 53 articles while Stata is used alone in only 14 -- but the use of R in conjunction with Stata is quite common, occurring in 26 different articles.  As expected, R used alone is the most common pattern observed in JASA, occurring in 127 papers from 2024. R and Python were used together in 22 papers, while R was used with C/C++ in 20 papers. Thus, here again we see a marked difference between statistics -- where R remains uniquely dominant -- and quantitative economics and political science, where both R and Stata are common and we observe a clear trend toward combining different types of software.

%We note that
Many of these patterns align with what one might have expected (economists tend to use Stata; statisticians favor R), but it's satisfying to have data to support those assumptions. Other trends, such as R and MATLAB now being equally common among economists, were less obvious ex ante. Another clear takeaway is the number of manuscripts, across the board, that utilize multiple software platforms. This may reflect increased collaboration across academic disciplines, or the growing ease with which researchers can learn and adopt multiple tools, thanks to the proliferation of online tutorials, open-source communities, and more accessible training resources.  It is also worth noting that some software packages rely on libraries implemented in other programming languages. For example, many R packages call underlying C or C++ code. Our process likely does not capture these dependencies, but this is a useful and important conversation to have with students.

Students also found something that our automated look at large repositories could not: the type of statistical software used was evident (either through reproduction code availability or through explicit text description) in 91 percent of articles examined in JASA, while this was true of only 71 percent of articles in the \emph{American Political Science Review}; the \emph{American Economic Review} fell between the two at 82 percent.  This pattern may mainly reflect the types of articles in the journals: theoretical papers or Nobel lectures might not require new statistical analysis.  We also investigated whether differences in journal reproducibility policies might contribute to this pattern. However, all three journals maintain substantial reproducibility and transparency requirements that are imposed during the manuscript review process.  % OO: changed "evaluated" to "imposed"

% Checked the comments in both the three tabs by the two special RAs, and in 2000 comment and ambiguity fields in the qualtrics entries.
% There were no mentions of LimDep
% There were no mentions of EViews
% There were no mentions of Minitab
% There were no mentions of RATS
% gretl and gephi come from the two RAs.
% two Gurobi entries come from the two RAs, depending on data clearning there may be one from qualtrics but simpler to exclude.
% 8 instances of "arcgis" or "qgis" from the two RAs, more from qualtrics 

\begin{figure}
\begin{center}
\includegraphics[width=0.8\textwidth]{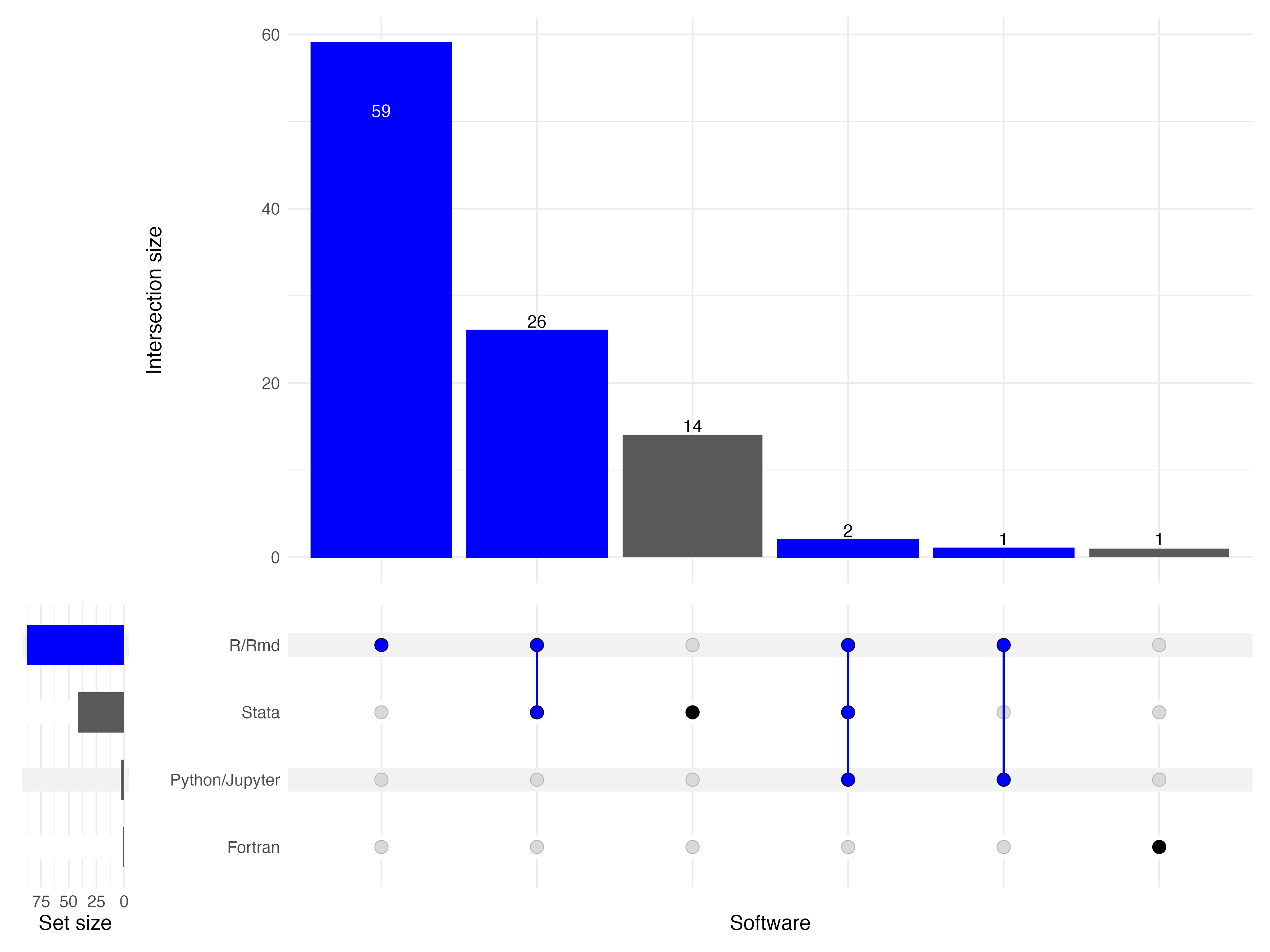}
\end{center}
\caption{Software in 2024 issues of the \emph{American Political Science Review}. Restricted to 103 articles (71 percent) in which software could be categorized out of 146 articles in total. \label{fig:2024apsr}}
\end{figure}

\begin{figure}
\begin{center}
\includegraphics[width=0.8\textwidth]{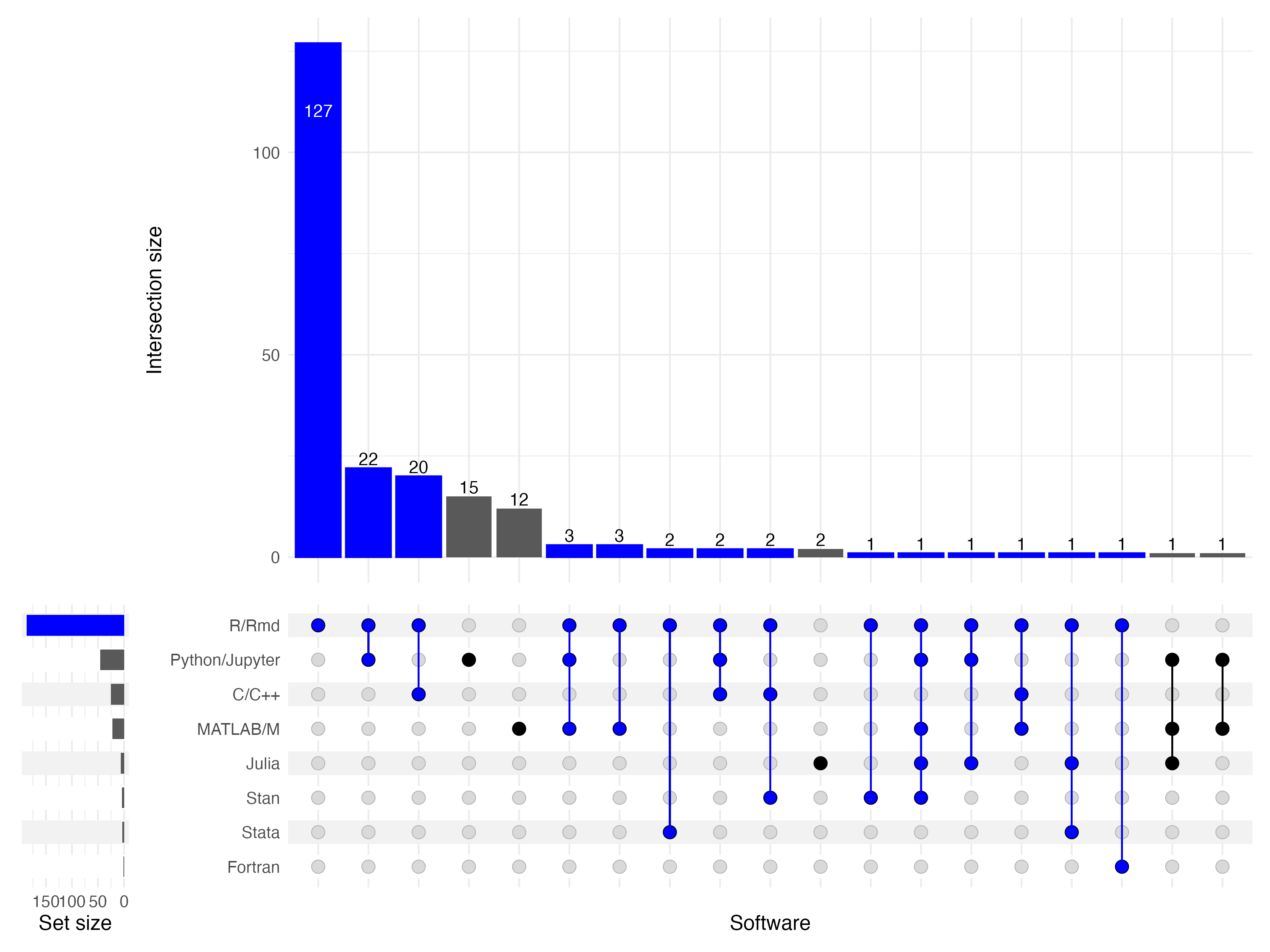}
\end{center}
\caption{Software in 2024 issues of the Journal of the American Statistical Association. Restricted to 218 articles (91 percent) in which software could be categorized out of 240 articles in total. \label{fig:2024jasa}}
\end{figure}

\section{Discussion}
\label{sec:conc}

The student activity was introduced as part of a broader, campus-wide data collection effort, with students contributing to a shared dataset. Prior to implementation, instructors expressed some concern that students might view the task as busy work, particularly because it involved navigating journal folders rather than traditional problem solving. In practice, the opposite occurred. Students expressed excitement about participating in a collaborative data collection effort and reported little to no prior awareness that academic manuscripts are often accompanied by online repositories containing data, code, and documentation; this aligns with \citet{Ostblom2022JSDSEopinionated}, who observe that undergraduate students are often unaware of the reproducible and auditable processes underlying data-centric research.

Furthermore, in our experience, students found exploring journal websites and academic articles both fun and eye-opening. The comments left by the students on the survey provide evidence the activity was well-designed: \\
\indent \textit{``I didn't encounter any issues during this task. All the DOIs that were assigned to me were easy to navigate. I was able to find the reproduction materials pretty quickly."} \\
\indent \textit{``Was confused at first how the websites work / where to find the data but figured it out and it became quick!"}

General consensus was that JASA articles were the most variable and difficult to navigate, but the students enjoyed browsing the articles and learned to differentiate between common sections - derivations, simulations, and data analyses.  Similarly, the students often pointed out when a manuscript did not meet their reproducibility criteria:
\\
\indent \textit{``mentions github repository with more details but could not find link"} \\
\indent \textit{``there seems to at one point have been a reproduction package but it is currently temporarily unavailable"}

Student engagement was evident in a senior-level statistics seminar, where there was audible cheering upon seeing that R was gaining popularity in political science. Meanwhile, some students questioned Python’s relatively low usage in JASA articles. This prompted a conversation about the scope of our conclusions, including how journal focus and subfield concentration mean that JASA does not represent statistics as a whole.  

We also shared these graphics with both an intermediate econometrics and an advanced microeconometrics course in the economics department.  Sharing the data in econometrics courses served different purposes at different moments in the courses.  It motivated training in Stata so that students could engage with the most common language in reproduction packages; it also motivated advanced assignments that took place in multiple languages (Stata and R) or that required translation (from Stata to Python or R) since our data show multilingual research environments to be increasingly common.  These applications provide one way to situate computational instruction within authentic disciplinary practice, as advocated by \citet{deveaux}, while our findings provide empirical context for decisions about which software environments students are likely to encounter. 
%partly to motivate the decision to train students in these languages, and to motivate the use of give students the choice of software (Stata, R, or Python) for their lab assignments. Interestingly, some students, after seeing these trends, chose to complete their labs in multiple software environments to broaden their skills.

This activity described in section \ref{sec:assignment} is intended as a first step in introducing students to the reproducibility pipeline.  To build on the initial data-collection phase, we have identified several follow-up activities that reinforce key learning objectives—particularly around reproducibility and software fluency. Some of these tasks we have implemented ourselves, while others serve as recommendations for future courses depending on the pedagogical goals:

1.	 Provide students with the full dataset collected and have them create their own data visualizations that summarize the current software landscape.  The possibilities here are numerous, but the nature of the data (time series/categorical variables) makes this an interesting exercise in data visualization.  

2.	Ask students to pick a manuscript that includes reproduction files, reproduce its primary analysis, and then translate the original code to another software environment. Comparing the two versions helps students appreciate both the similarities and subtleties in different programming languages.  

3.	Reproduce and expand upon an existing data analysis from the collection. Students should identify and evaluate the authors’ data-processing choices and note how they might do things differently. If the original analysis lacks explicit assumption checks (e.g., residual plots), students should perform and interpret them as part of this extension.  

4.	Have students assess the reproducibility of the provided code and documentation. Is the code well-commented and easy to follow? Does a README file clearly introduce the project and explain how to run everything? When present, do automated testing suites or example scripts provide additional evidence that the code behaves as intended? These questions help students see the real-world importance of sharing reproducible research.

5. For students particularly interested in the activity, offer research assistant opportunities in which they could gather more data. This could focus either on specific journals to create a more comprehensive picture for particular subfields or on disciplines we have not yet considered. 

\section{Conclusion}
\label{sec:conc2}

The activity provides a low-cost way to expose students to reproducibility as it is practiced in published research. Straightforward instructions allowed students to navigate reproduction materials while forming their own judgments about what makes research easy or difficult to reproduce. This approach is consistent with calls to make reproducible and auditable workflows an explicit part of data-centric education \citep{rethlefsen2022interdisciplinary} and complements other classroom activities designed to engage students directly with reproducible research practices \citep{gligoric2024class,liu2022promoting, ballyes}.  Anecdotally, students expressed an appreciation for having a better understanding of how academic research is organized, accessible, and presented across different fields.

A key takeaway from this activity is the importance of being multilingual, however we recognize that limited class time (and student interest) can prevent a deep dive into multiple software environments in any single course. In most of our classes, software is primarily a means to carry out analyses and we cannot devote too much time, if any, to learning more than one tool. Still, even brief exposure to additional languages can be valuable: students may be proficient in Stata, for instance, yet they should realize that the coding principles they’ve learned will map onto R, Python, or other platforms. For introductory courses, perhaps a single class should focus on a single language.  But  students should know that there is more than one of them.  

Even in statistics, we can see that though R is by far the most popular language, the use of multiple languages is not uncommon. The present manuscript is no exception.  In writing this paper, we used both R and Stata for web scraping; one author used Stata for assigning students to tasks, and two others used R for producing final visualizations.

\section{Limitations}
Several limitations have been noted throughout the manuscript, but it is worth summarizing them here.  First, our conclusions are based on a selected set of journals and repositories and should not be interpreted as representative of all research conducted in economics, political science, or statistics. Second, our student survey focused on a predefined set of common software tools.  Less common software and packages were occasionally mentioned in student comments, but we cannot be sure how frequently they are mentioned because we did not ask students to systematically search for them.  For example, there were two student mentions of Gurobi, one of gretl, and one of Gephi.  Geographic analysis packages such as ArcGIS and QGIS were mentioned in a number of instances as well.  Third, the automated and manual data collection approaches each have limitations: repository metadata do not always fully describe the contents of reproduction packages, while manual classifications are subject to occasional student error. Fourth, in the original design of the student activity, we used the terms reproducibility and replicability interchangeably, which may have caused confusion among students about the distinction between reproducing results using existing data and code and replicating findings using new data.  This experience has led us to be more explicit and intentional in distinguishing between reproducibility and replicability in our own teaching and research.  Finally, because the data sources differ in coverage and availability across disciplines, comparisons should be interpreted as descriptive rather than definitive measures of software use within entire fields.

\section*{Data Availability}
The data collected by the students and authors that support the findings of this study are available in an OSF repository: \url{https://osf.io/72ajq/}. 

\section*{AI Use Statement}
No generative artificial intelligence tools were used in the creation of the student assignment or the code for data scraping, analysis, and figure creation.  Similarly, no AI tools were used in the substantive writing of this manuscript. Limited use of ChatGPT based on GPT-5.5 was made for language editing and polishing.

\section*{Acknowledgments}
The authors are grateful to Richard De Veaux for helpful comments on an early draft; to David Zimmerman for crucial early stage work in the project and for providing his support throughout; and to Isabel Beckrich and Jonathan Hartanto for excellent research assistance.

\section*{Ethics Approval and Consent}
This study was reviewed by the Williams College Institutional Review Board and determined to be exempt from IRB review, as it presents no risk to participants and relies exclusively on publicly available data.   The IRB exemption approval number is 2025-23, and Williams College IRB registration number is 00003239.  

\bigskip
\begin{center}
{\large\bf SUPPLEMENTARY MATERIAL}
\end{center}

\begin{description}

\item Data details, sample instructions, and a survey template are provided in the online appendix.

%\item[Title:] Brief description: sample student activity prompt. (file type)

\end{description}

%\section{BibTeX}

\bibliographystyle{asa-jsdse}
\bibliography{socialsciencesoftware_new}

\end{document}